\documentclass[conference]{IEEEtran}
\IEEEoverridecommandlockouts
\usepackage{cite}
\usepackage{amsmath,amssymb,amsfonts}
\usepackage{algorithm}
\usepackage{algpseudocode}
\usepackage{graphicx}
\usepackage{textcomp}
\usepackage{xcolor}
\usepackage{multirow}
\def\BibTeX{{\rm B\kern-.05em{\sc i\kern-.025em b}\kern-.08em
    T\kern-.1667em\lower.7ex\hbox{E}\kern-.125emX}}

\usepackage{hyperref}
\usepackage{tikz}
\begin{document}
\title{An Improved PMOS-Based Low Dropout Regulator Design for Large Loads}

%
%

\author{
\IEEEauthorblockN{Arijit Saha}
\IEEEauthorblockA{\textit{Dept. of Electronics and Tele-communication Engineering} \\
\textit{Jadavpur University}\\
Kolkata – 700 032, INDIA \\
arijitfeb01@gmail.com}
\and
\IEEEauthorblockN{Ayan Biswas}
\IEEEauthorblockA{\textit{Dept. of Electronics and Tele-communication Engineering} \\
\textit{Jadavpur University}\\
Kolkata – 700 032, INDIA \\
 ayanbiswas@ieee.org}
\and
\IEEEauthorblockN{Supriya Dhabal}
\IEEEauthorblockA{\textit{Electronics \& Communication Engineering Department} \\
\textit{Netaji Subhash Engineering College}\\
Kolkata - 700152, INDIA \\
supriya\_dhabal@yahoo.co.in }
\and
\IEEEauthorblockN{Palaniandavar Venkateswaran}
\IEEEauthorblockA{\textit{Dept. of Electronics and Tele-communication Engineering} \\
\textit{Jadavpur University}\\
Kolkata – 700 032, INDIA \\
pvwn@ieee.org}
}

\maketitle

\begin{abstract}
A stable low dropout (LDO) voltage regulator topology is presented in this paper. LDOs are linear voltage regulators that do not produce ripples in the DC voltage. Despite the close proximity of the supply input voltage to the output, this regulator will maintain the desired output voltage. Based on a detailed comparison between NMOS and PMOS-based LDOs, we decided to opt for a PMOS design because it does not require an additional charge pump as compared to NMOS. A demonstration of how Miller capacitance enhances overall design stability is also presented here. Multiple pass elements are arranged in parallel in order to increase the current carrying capacity of the pass network.\\
\end{abstract}

\begin{IEEEkeywords}
Low Dropout Regulator, Operational Trans-conductance Amplifier, Miller Compensation, Programmable Output Voltage

\end{IEEEkeywords}
\section{Introduction}

Since the development of the Very Large Scale Integration (VLSI) industry and the consequent scaling down of devices following Moore's law, transistor sizes have significantly decreased. As a result, supply voltages have also been reduced. Therefore, voltage regulators are needed, which output a fixed voltage despite varying input voltages. The Integrated Circuit's (IC's) other parts receive this fixed output voltage as a supply. There are two main types of regulators: linear and switching. Low dropouts (LDOs) are linear regulators that are commonly used in VLSI chips. The reduction in supply voltage due to scaling has made LDOs an essential component of power management ICs because we require lower input-output voltage differences, i.e., lower dropout. 

An LDO consists of four main blocks: an error amplifier, a pass element, a feedback network, and a load. An error amplifier is a differential amplifier based on an Operational Trans-conductance Amplifier (OTA). Like an Operational Amplifier (OPAMP), it has similar characteristics. On the other hand, OTAs are designed for capacitive loads, while OPAMPs are designed for resistive loads.  Metal Oxide Semiconductor Field Effect Transistors (MOSFETs) or Bipolar Junction Transistors (BJTs) can be used as the pass element. It is preferable to use MOSFETs because they are not very sensitive to temperature changes. A simple resistive voltage divider network can be used as a feedback network.

The remainder of this paper is organized as follows.
Section~\ref{sec:lit_review} reviews the related works in this field.Section~\ref{sec:pmos_nmos} reviews the technical differences between p-channel Metal Oxide Semiconductor (PMOS) and n-channel Metal Oxide Semiconductor (NMOS)-based LDOs. Section~\ref{sec:error_amp} presents the design of the error amplifier for the LDO. The significance of the miller capacitance has been discussed in Section~\ref{sec:miller_cap}. Lastly, the final design of the PMOS-based LDO~\footnote{The design has been simulated using LTspice\cite{ref_lt_spice} developed by Linear Technology Corporation and Analog Devices, Inc. It is widely used for SPICE-based~\cite{spice} simulations.} has been discussed in Section~\ref{sec:pmos_ldo}.

\section{Related Works}\label{sec:lit_review}

This section of this paper aims to explore recent developments in the field of LDO linear regulators, specifically focusing on designs that improve the efficiency and stability of these regulators while maintaining low output impedance. Recent developments in this field are discussed below:

 The authors in ~\cite{choi} described a design for an LDO linear regulator with an ultra-low-output impedance buffer in their work. The authors propose a new architecture that reduces the output impedance of the LDO linear regulator by using an additional buffer circuit, which improves the regulator's load transient response and stability. Through simulations and measurements, the authors show that their proposed design enhances the efficiency and stability of the LDO linear regulator while maintaining low output impedance. 
 
 The researchers in ~\cite{alica} have presented a design for a low voltage, low dropout (LDO) regulator in their article with the ability to adjust to two different output voltages (1V or 1.8V). The authors proposed a new architecture that utilizes 0.35 micron CMOS technology and is able to drive a load of up to 50mA while maintaining a maximum dropout voltage of only 200mV. The LDO design is optimized to minimize the quiescent current and to extend the battery life of portable devices. 
 
 In a recently developed pioneering work ~\cite{razavi}, the importance of designing LDO regulators specifically for the circuits they will be powering in mixed-signal systems has been explicitly discussed. The article highlights the role of LDOs in isolating circuits from noise and the need for tailored designs for optimal performance. The author also notes the difference in design for an LDO serving a flash analog-to-digital (ADC) converter versus one serving a VCO. 
 
 The authors of ~\cite{luo} proposed a new NMOS-LDO voltage regulator in their paper with a fast transient response. The design is a dual-loop circuit, with an inner loop that increases the regulating speed of the output voltage. The authors also use an NPN to move the dominant pole from the output point to the inner of the LDO chip, eliminating the need for a miller capacitor or other on-chip compensated capacitor. The simulation results in the 0.35$\mu$m CMOS process show that the rise time and fall time of LDO is about 2.11$\mu$s and 4.26$\mu$s for 10$\mu$A ~10mA step change of load current. Another NMOS-based LDO design has been presented by the researchers in their article ~\cite{Giustolisi2008}. They have proposed an NMOS low drop-out voltage regulator for on-chip power management. The circuit is internally compensated and does not require any external components, and it uses a dynamic biasing strategy and a clock booster to drive the NMOS power transistor in a power-efficient fashion without limiting the speed response of the regulator. Their simulation results confirm the effectiveness of the proposed approach.

\section{Comparative study between performances of PMOS and NMOS LDOs} \label{sec:pmos_nmos}

\begin{figure}[ht]
\includegraphics[width=\linewidth]{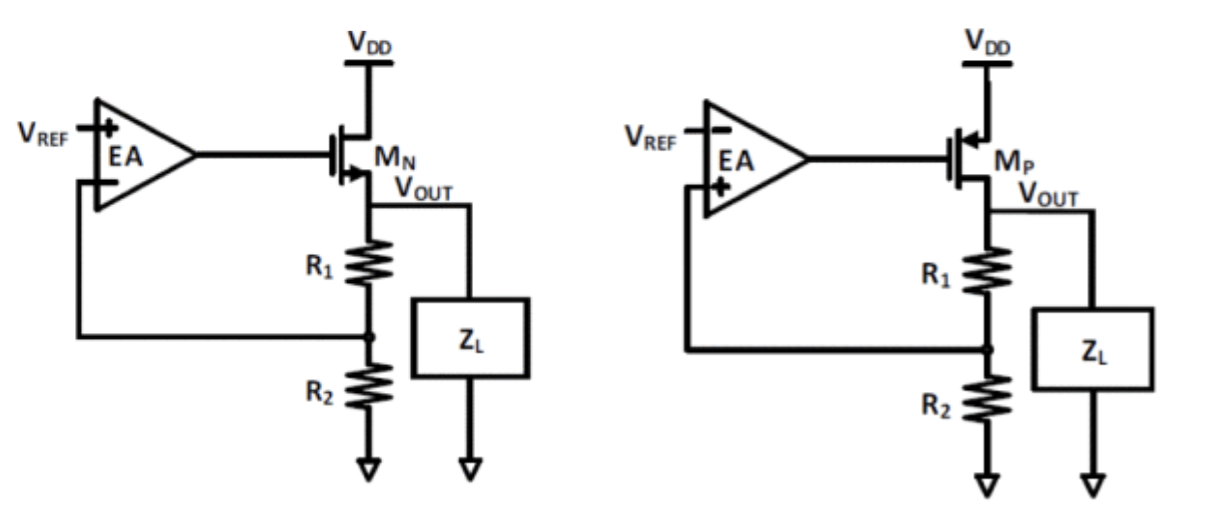}
\caption{Block level design of PMOS and NMOS based LDO} \label{fig1}
\end{figure}

The two main architectural types of the LDO~\cite{ref_article12} is shown in Fig.~\ref{fig1}, the difference between them is the type of pass transistor. The first one is a PMOS pass transistor, whereas the second is an NMOS pass transistor.

Three major components make up a typical LDO circuit, namely, a high gain error amplifier ~\cite{ref_article12}, a pass transistor ~\cite{pass_tran}, and a feedback network. Using the pass transistor to control the load current, the High gain error amplifier compares output voltages with reference voltages, and the error amplifier receives a return voltage signal from resistors that act as voltage-voltage feedback to sense output voltages from the LDO.

\subsection{The reason behind selecting a PMOS-based design}

The dropout voltage of PMOS is lower than that of NMOS. However, as a common drain, the NMOS pass transistor is connected, which leads to a small output resistance at high load currents due to the increase in trans-conductance. This makes it more cumbersome to fabricate the IC since an additional charge pump ~\cite{ref_article3} is required to support a wide range of load currents. PMOS LDO has higher loop gain as compared to NMOS-based LDO. PMOS is, however, a bit slower compared to NMOS since the mobility of electrons is greater than that of holes, thus PMOS design occupies a larger area which accounts for larger capacitances. This makes the PMOS LDO ~\cite{ref_article5} slower than its NMOS counterpart. Thus, we have selected PMOS for our design by balancing odds and favors.

\section{Error Amplifier Design} \label{sec:error_amp}

The error amplifier~\cite{ref_article4} is a two-stage OTA. The first stage is the differential amplifier stage. The second stage is a gain-enhancing common source stage. We have used a current mirror biasing for the first stage formed by MOSFETs $M_1$ and $M_6$. The MOSFETs $M_2$ and $M_3$ form the inverting and non-inverting terminals of the OTA respectively. The design has been done for 180nm technology. The current flowing through $M_6$ is copied in $M_1$ and is twice the current flowing through $M_4$ and $M_5$ each since the same current flows through $M_4$ and $M_5$ due to equal gate-source voltage. $C_c$ is the Miller Compensation capacitance and $C_L$ is the load capacitance. The miller capacitance has been added to increase the stability of the error amplifier.  

\begin{figure}[ht]
\includegraphics[width=\linewidth]{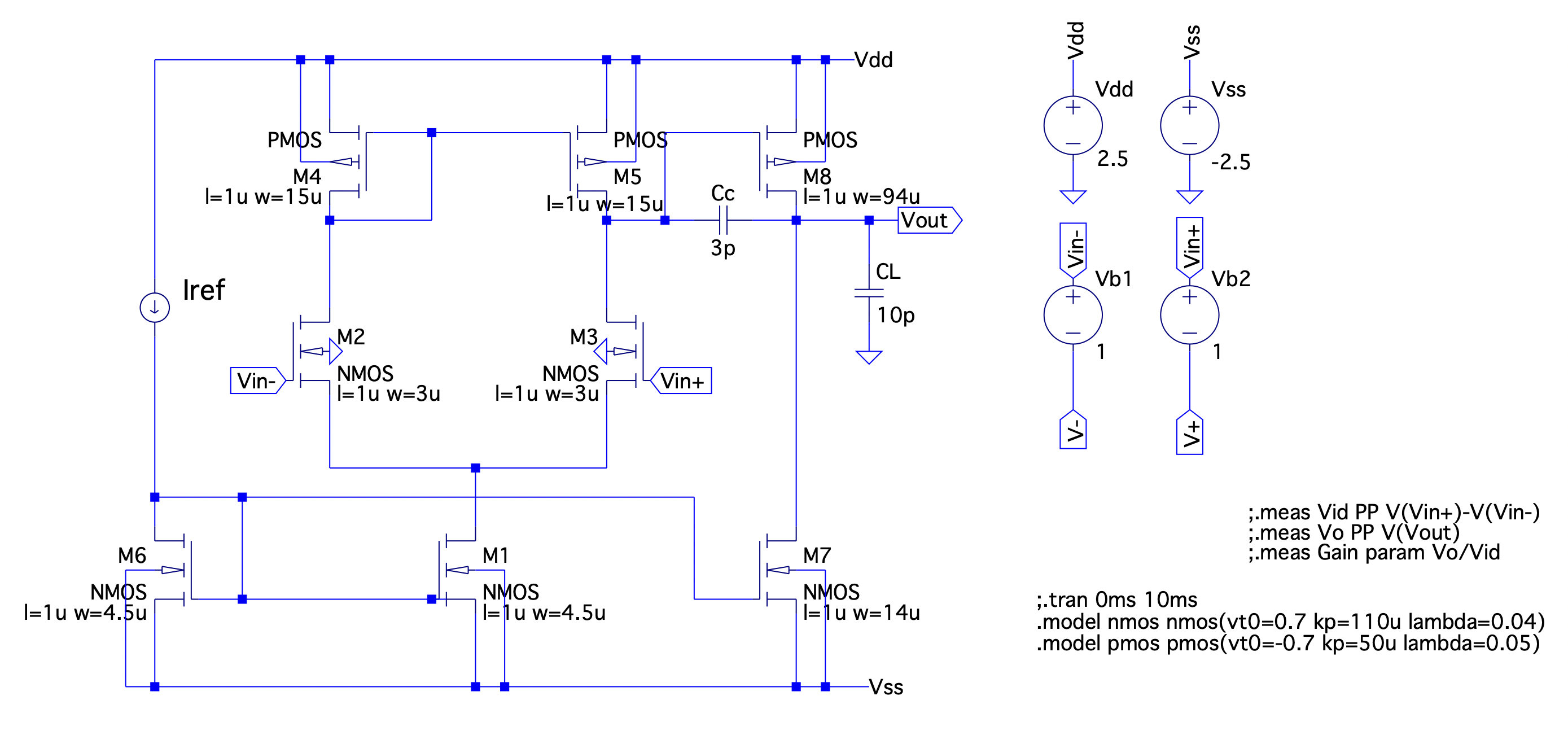}
\caption{Circuit diagram of error amplifier} \label{fig2}
\end{figure}

\section{Significance of Miller Compensation Capacitance} \label{sec:miller_cap}

The compensation capacitance has been used to stabilize the system by increasing its phase margin. MOSFET $M_4$ is diode-connected, so it has a very low output impedance ($~1/g_{m_{4}}$). Hence, the overall port impedance at the drain of $M_4$ is low. However, the port impedances at the drains of $M_5$ and $M_8$ are very high (comparable to the $r_o$). This forms two low-frequency poles ($~1/(r_{o}*C_{L})$). Each pole contributes a $-90^o$ phase shift, and thus a total phase shift of $-180^o$ at low frequencies. This results in $0^o$ phase margin and the OTA will oscillate in negative feedback. Thus, we are adding a compensation capacitor~\cite{ref_article2} between the drain of $M_5$ and $M_8$. Now, the output impedance at the drain of $M_5$ will see a much larger capacitance according to Miller’s theorem and this pole will shift towards a lower frequency. The other pole shifts towards higher frequency. Thus the system will essentially become a 1st order system with a significant phase margin ($~60^o$). Thus, the OTA will function as an amplifier instead of an oscillator.

\section{Proposed PMOS LDO} \label{sec:pmos_ldo}

The architecture of this LDO is similar to a basic LDO regulator. A voltage reference of 1.2V~\cite{ref_article6} is generated by the bandgap which is given to the negative terminal of the error amplifier. The output of the error amplifier block is fed to the gate terminal of the PMOS pass network and at the drain, a resistive divider network is connected. The feedback voltage from the resistive divider is fed back to the positive terminal of the error amplifier to ensure negative feedback. 

\begin{figure}[ht]
\includegraphics[width=\linewidth]{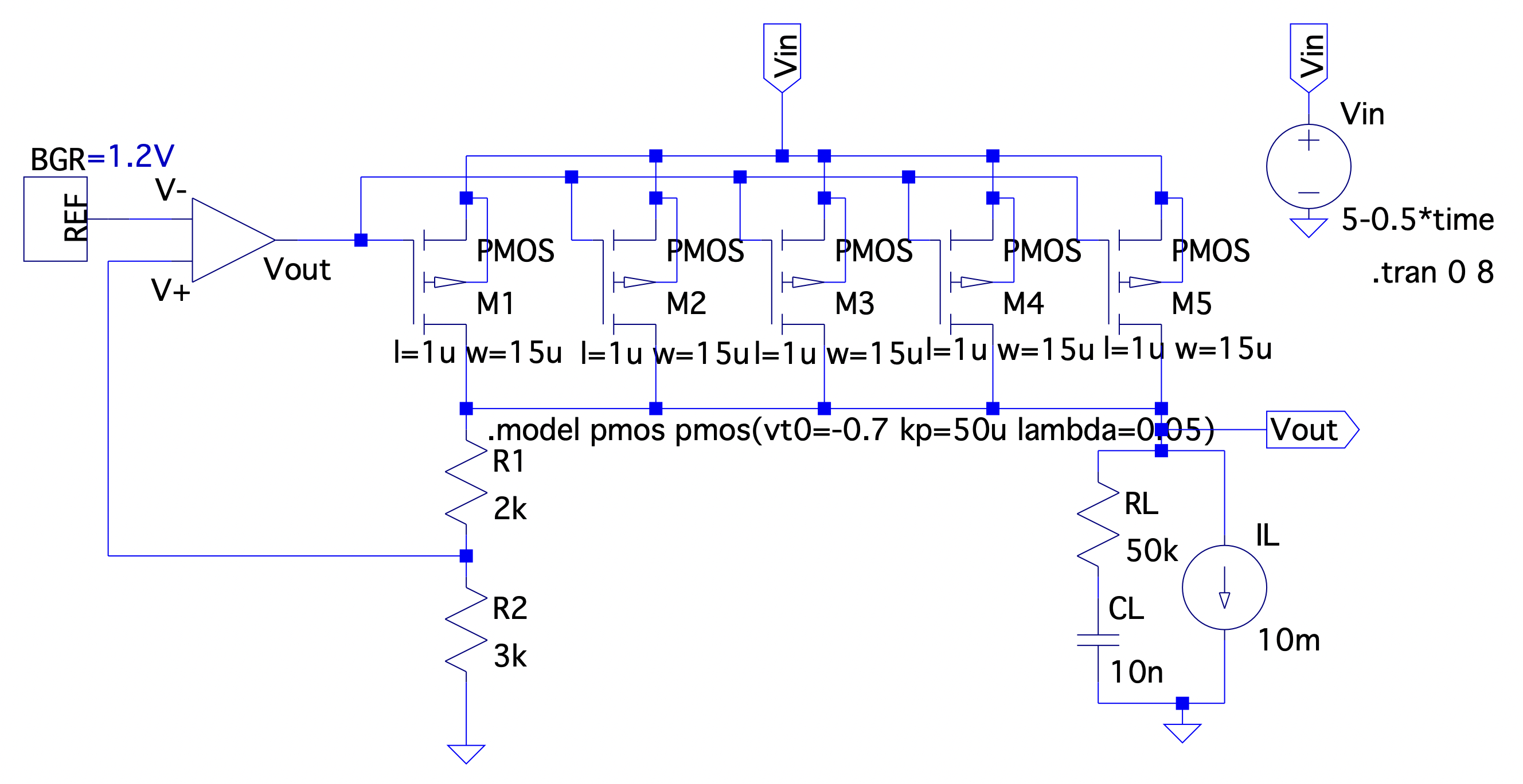}
\caption{Circuit design of PMOS based LDO} \label{fig3}
\end{figure}

\subsection{Working Principle of Circuit}

From Fig.~\ref{fig2}, the error amplifier is in negative feedback. This is because according to Barkhausen criteria, the total phase shift should be $-180^o$ for negative feedback. The gate-drain phase shift is $-180^o$ and the voltage is fed back to the positive terminal. Therefore, the total phase shift is calculated to be $-180^o$. We can assume a virtual short condition for negative feedback in OPAMP. Thus, $V_+ = V_- = V_{REF}$ which leads to the following working formula for $V_{OUT}$.

\begin{equation}
V_{out} = V_{REF} * (R_{1} + R_{2})/R_{2}
\end{equation}

Now, let's understand the LDO regulation principle intuitively. If the load current increases, the current through the pass element cannot increase immediately. It will undergo some transient. Initially, the required additional load current is drawn from the load capacitor. This will lead to a decrease in the output node voltage. This output voltage reduction will lead to a decrease in the feedback voltage. Therefore, the gate voltage of the pass element will also decrease, because the voltage has been fed back to the positive terminal. Thus, the source-gate voltage of the PMOS pass element increases, which finally increases the current through the pass network to the required level. The same is the scenario when the load current drops. In this way, the output voltage and the load current are maintained by the LDO.

\subsection{Simulation and Analysis}

\begin{figure}[ht]
\includegraphics[width=\linewidth]{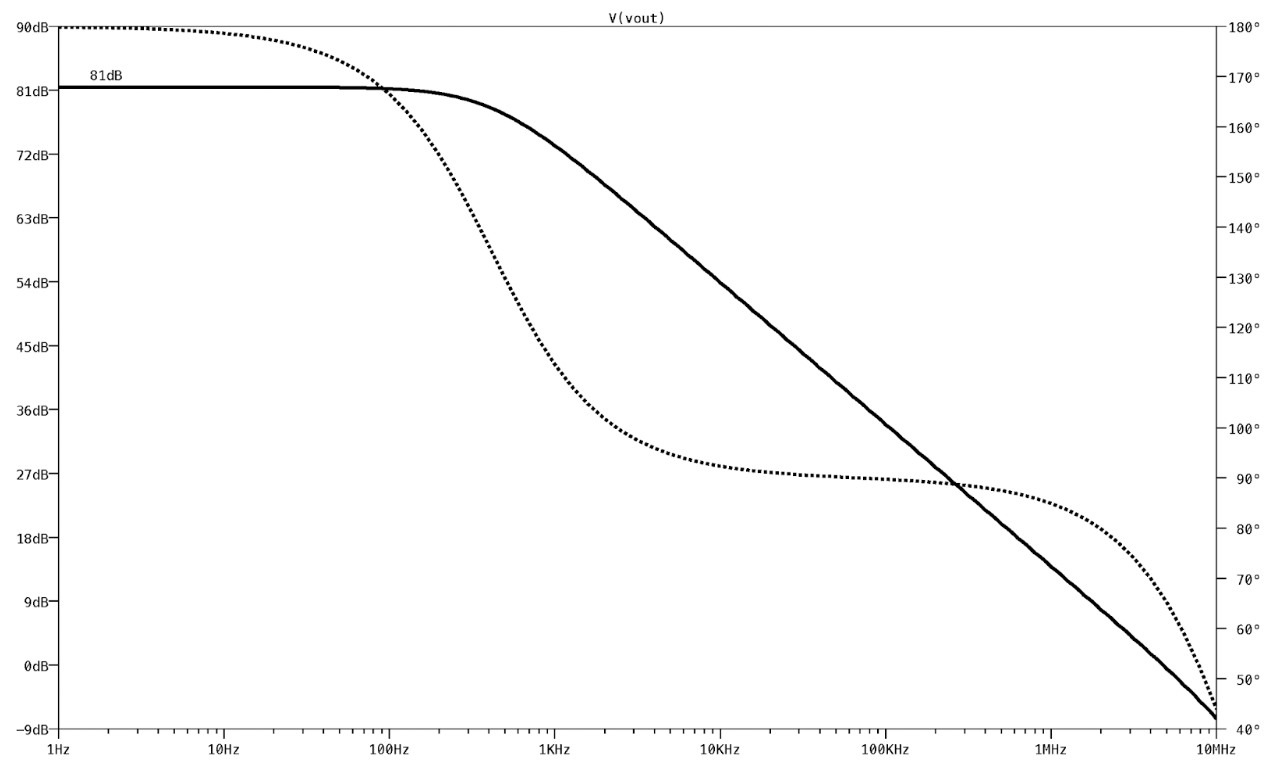}
\caption{Bode plot of the OTA} \label{fig4}
\end{figure}

The 2-stage OTA ~\cite{ref_article7} forming the error amplifier was simulated in LTspice. The OTA was designed for a DC differential voltage gain greater than 5000 and a Gain Bandwidth Product (GBP) of 5MHz. The bode plot of the OTA was obtained as shown in Fig. ~\ref{fig4}.

\begin{table}[ht]
\caption{Simulation and analysis results of the PMOS based LDO}\label{tab1}
\begin{tabular}{|l|ll|}
\hline
Parameters                       & \multicolumn{2}{l|}{Values}                  \\ \hline
Input Voltage                    & \multicolumn{2}{l|}{1.0V - 5.0V}             \\ \hline
Output Voltage                   & \multicolumn{2}{l|}{2.0V}                    \\ \hline
Reference Voltage                & \multicolumn{2}{l|}{1.2V}                    \\ \hline

\multirow{3}{*}{Dropout Voltage (variation with $I_{LOAD}$)} & \multicolumn{1}{l|}{$I_{LOAD}$  = 5mA}  & 0.3V    \\ \cline{2-3} 
                                 & \multicolumn{1}{l|}{$I_{LOAD}$ = 7.5mA} & 0.45V   \\ \cline{2-3} 
                                 & \multicolumn{1}{l|}{$I_{LOAD}$ = 10mA}  & 0.6V    \\ \hline
\multirow{3}{*}{Power Consumed (variation with $I_{LOAD}$)}  & \multicolumn{1}{l|}{$I_{LOAD}$ = 5mA}   & 7.42mW  \\ \cline{2-3} 
                                 & \multicolumn{1}{l|}{$I_{LOAD}$ = 7.5mA} & 10.25mW \\ \cline{2-3} 
                                 & \multicolumn{1}{l|}{$I_{LOAD}$ = 10mA}  & 12.31mW \\ \hline
Miller Capacitance               & \multicolumn{2}{l|}{3pF}                     \\ \hline
Maximum Tolerable Load Current   & \multicolumn{2}{l|}{23mA}                    \\ \hline
\end{tabular}
\end{table}

The DC gain has been found to be 81dB = 11220 which is greater than 5000 as expected. The 3dB cutoff frequency is found to be 440Hz. Thus, the GBP evaluates to 5MHz which matches our design constraints.

\begin{figure}[ht]
\includegraphics[width=\linewidth]{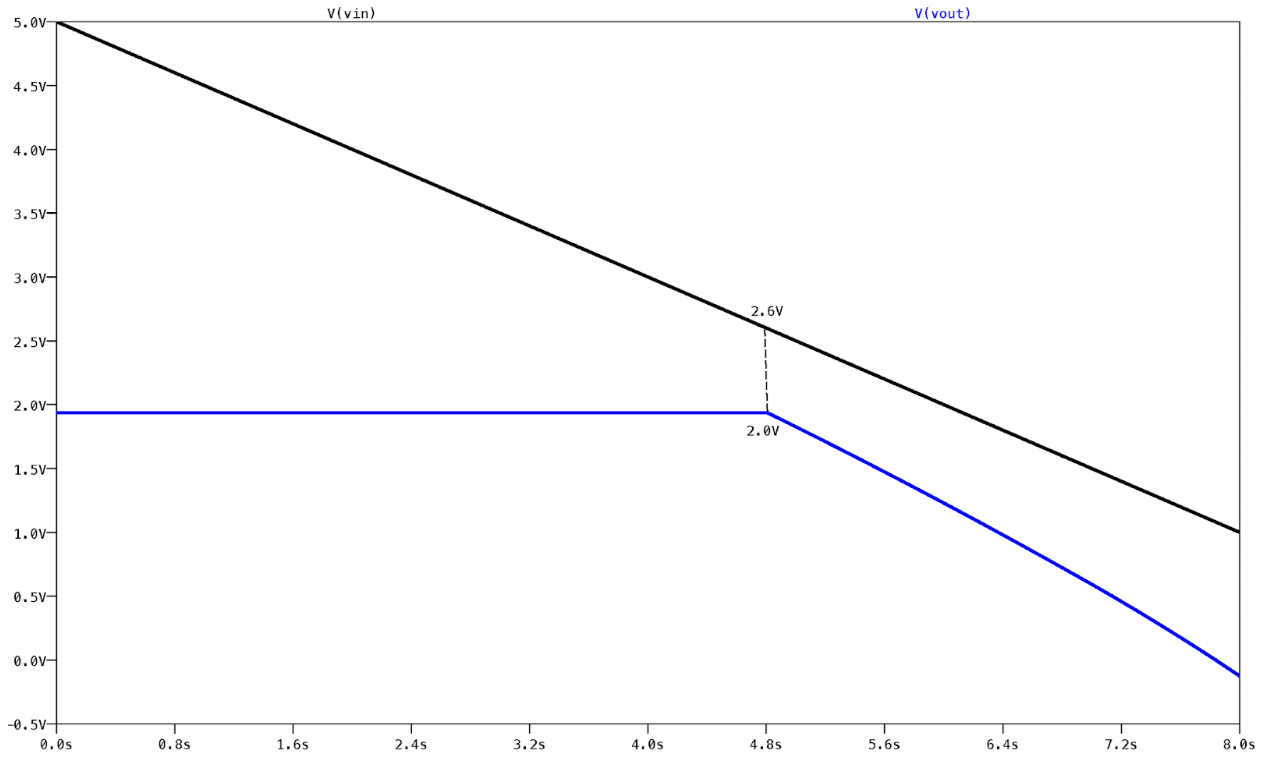}
\caption{The LDO circuit was simulated for a load current of 10mA and the input supply was varied from 5.0V to 1.0V. The output was maintained at 2.0V as given by equation (1) until $V_{in}$ drops below 2.6V. Therefore, the dropout voltage is (2.6 - 2.0) V = 0.6V} \label{fig5}
\end{figure}

\begin{figure}[ht]
\includegraphics[width=\linewidth]{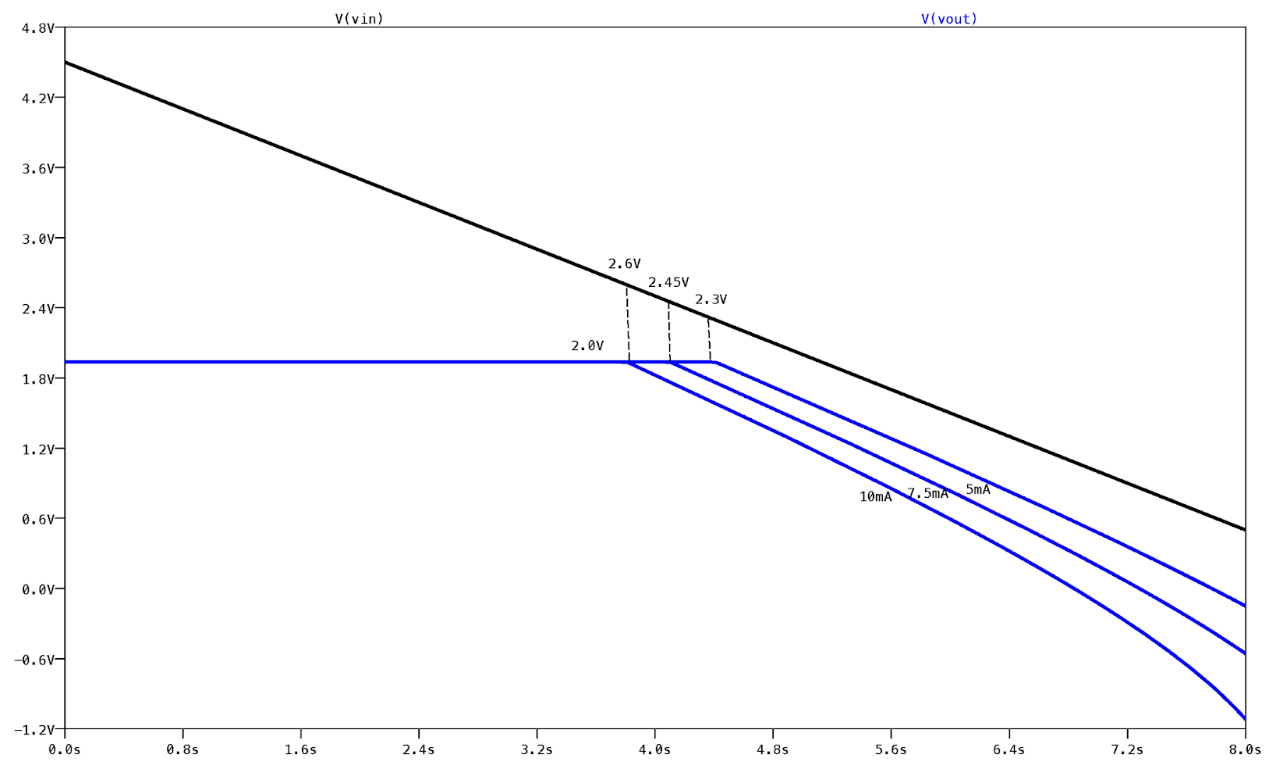}
\caption{Behaviour of the LDO for varying load currents} \label{fig6}
\end{figure}

We observed the behavior of the LDO for varying load currents in Fig.~\ref{fig6}. It is found that as load current increases the performance of the LDO degrades since the dropout voltage increases~\cite{ref_article8, ref_article13}. The LDO supplies a maximum current of 23mA after which output voltage cannot be regulated. We have also summarized the results of our design in Table~\ref{tab1}.

\section{Conclusion} \label{sec:conclusion}

This paper illustrates how low dropout (LDO) voltage regulator topology can be applied to voltage regulator design and why PMOS-based designs are preferred. The simulation clearly shows that the proposed technique can tolerate up to 23mA of current, which is an excellent result as compared to other LDOs. The PMOS-based design was stabilized with an efficient compensation technique and a Miller Capacitance of 3pF was used to achieve $~60^o$ phase margin.



\bibliographystyle{ieeetr}
\bibliography{citation}

\end{document}